\documentclass[aps,reprint,prx,amsfonts,amsmath,superscriptaddress]{revtex4-2}

\usepackage{graphicx}
\usepackage{times}
\usepackage{hyperref}
\usepackage{amsfonts}
\usepackage[caption=false]{subfig}
\usepackage[T1]{fontenc}

\begin{document}

\title{Linear-time classical approximate optimization of cubic-lattice classical spin glasses}

\author{Adil A. Gangat}
\email{adil.gangat@ntt-research.com}
\affiliation{Physics \& Informatics Laboratories, NTT Research, Inc., Sunnyvale, CA 94085}
\affiliation{Division of Chemistry and Chemical Engineering, California Institute of Technology, Pasadena, CA 91125}
\date{\today}
\begin{abstract}
Demonstrating quantum speedup for approximate optimization of classical spin glasses is of current interest. Such a demonstration must be done with respect to the best-known scaling of classical heuristics at a given optimality gap of a given problem.  For cubic-lattice classical Ising spin glasses, recent theoretical and experimental developments open the possibility of showing quantum speedup for approximate optimization with quantum annealing.  It is therefore desirable to understand the optimality-gap range over which such a speedup should be searched for.  Here we show that on cubic-lattice tile-planting models, classical meta-heuristics that are linear-time by construction can reach optimality gaps at which simulated annealing and parallel tempering exhibit super-linear scaling.  This implies that the optimality gaps achieved by linear-time classical meta-heuristics can serve as useful upper bounds for the optimality-gap range over which quantum speedups in approximate optimization should be searched for.  We also explain how classical heuristics with fixed scaling that is beyond-cubic can provide upper bounds to optimality-gap ranges for beyond-quadratic quantum speedups in approximate optimization.  These results encourage the development of classical heuristics with fixed scaling that achieve optimality gaps as small as possible.
\end{abstract}
\maketitle

\section{Introduction}
\label{sec:intro}

Discrete combinatorial optimization is an area where quantum computing is hoped to one day have a performance advantage over classical computing, and finding ways to realize this hope is an active area of research \cite{abbas2024challenges}.
Typically the discrete combinatorial optimization problem is mapped onto a classical spin-glass model such that the Hamiltonian encodes a cost function to be minimized and the ground state corresponds to the optimal solution of the optimization problem.  

The simplest possible class of such Hamiltonians is the Ising spin glass in zero field, given by
\begin{equation}
H = \sum_{i,j} J_{ij}\sigma_i\sigma_j,
\label{eqn:Ham}
\end{equation}
where $J_{ij}\in\mathbb{R}$, $\sigma_i\in\{\pm1\}$, and $i$ and $j$ are site indices. A structured choice of $J_{ij}$ turns the Hamiltonian into a cost function of an NP-hard discrete combinatorial optimization problem \cite{barahona1982computational}.  Demonstrating a performance advantage for quantum computing over classical computing in computing either the groundstates (i.e., exact optimization) or low-lying excited states (i.e., approximate optimization) of Eq. (\ref{eqn:Ham}) would be an important milestone.

\subsection{Quantum speedup for approximate optimization}
The reason why \textit{approximate} optimization of spin-glass Hamiltonians is of interest in addition to exact optimization is that merely reduced-cost solutions can be of practical benefit in industrial contexts, and what is often desired in such contexts is a set of low-energy states that satisfy various constraints \cite{caracciolo, wangwei}.  Also, the outputs of approximate optimization algorithms could potentially serve as \textit{warm starts} (i.e., initial conditions that result in a shorter runtime compared to random initial conditions) for other approximate optimization algorithms or exact optimization algorithms.

There is a long history of investigating the possibility of quantum advantage (i.e., some type of performance advantage of quantum algorithms over classical algorithms) for the optimization of classical spin glasses \cite{mohseni2022ising, abbas2024challenges}. A key performance metric is the time-to-solution (TTS) in the case of exact optimization, or time-to-epsilon (TT$\varepsilon$) in the case of approximate optimization. The TT$\varepsilon$ is defined as the time required to produce a solution \textit{at or below} an optimality gap of
\begin{equation}
\varepsilon=(E-E_{gs})/|E_{gs}|,
\label{eqn:error}
\end{equation}
where $E$ is the Hamiltonian energy of the computed configuration and $E_{gs}$ is the ground state energy. 

In this work we are concerned specifically with approximate optimization and the type of quantum advantage that is termed \textit{quantum speedup} \cite{ronnow2014defining}, which means the existence of a quantum algorithm with a smaller time complexity compared to the best-known time complexity of classical algorithms for a given problem and a given value of $\varepsilon$.  Ref. \cite{bauza2025scaling}, for example, claims that a particular form of quantum annealing achieves quantum speedup for a quasi-2D classical spin glass when $\varepsilon\gtrsim1\%$.  The time complexity of an algorithm is the scaling of its TTS or TT$\varepsilon$ as a function of system size ($N$).  An advantage in time complexity is sometimes referred to as a \textit{scaling advantage}.

For exact optimization there are a number of results showing the potential for quantum speedup \cite{farhi2001quantum, venturelli2015quantum, boulebnane2024solving, shaydulin2024evidence, sciorilli2025competitive}, though an actual demonstration remains absent.  Regarding quantum speedups in \textit{approximate} optimization, Refs. \cite{pirnay2024principle, jordan2025optimization} theoretically show the potential for it; Ref. \cite{bauza2025scaling} claims an actual demonstration of quantum speedup of D-Wave quantum annealing by comparing its time complexity to that of parallel tempering with isoenergetic cluster moves (a type of Markov-Chain-Monte-Carlo algorithm) for the approximate optimization of a class of quasi-2D spin glasses, but the more recent work of Ref. \cite{pawlowski2025closing} calls this into question by showing that the Simulated Bifurcation Machine \cite{goto2019combinatorial, goto2021high}, a classical heuristic inspired by nonlinear Hamiltonian dynamics, achieves similar or better time complexity than quantum annealing in the same class of problems.

It should be noted that demonstration of quantum speedup at \textit{any} $\varepsilon$ is of current interest, even if $\varepsilon$ is large enough to be reached in polynomial time by the quantum and classical heuristics under consideration.  For example, the claimed quantum speedup in Ref. \cite{bauza2025scaling} at $\varepsilon\approx1\%$ is where all the heuristics under consideration exhibit sub-cubic scaling.  The hope is that finding quantum speedups at such ``easy'' values of $\varepsilon$ will yield important fundamental insights that could in turn lead to quantum speedups of a more practically significant nature.

It would therefore be useful to have an efficient method for obtaining a nontrivial upper bound on the quantum-speedup search-space of $\varepsilon$.  Here we point out that classical heuristics that are \textit{constructed} to have a linear time-scaling can serve as such a method.  That is, the $\varepsilon$ obtained by performing approximate optimization with such a heuristic (call it $\varepsilon_{\textrm{lin}}$) serves as such an upper bound. The reason for this is that linear scaling is the best possible scaling for any optimization algorithm, quantum or classical; each spin must be assigned a value somehow, even if completely randomly, and since there are $N$ spins, there must
necessarily be $N$ spin assignments, which requires at least 
$\mathcal{O}(N)$ time to complete.  Trying to demonstrate quantum speedup at $\varepsilon>\varepsilon_{\textrm{lin}}$ is therefore futile because the linear-time approximate-optimization heuristic remains linear-time for $\varepsilon>\varepsilon_{\textrm{lin}}$ according to the definition of TT$\varepsilon$.

However, it is also intuitive that a finite but very large $\varepsilon_{\textrm{lin}}$ will not be \textit{useful} as an upper bound on the quantum-speedup search-space of $\varepsilon$.  We propose that a value of $\varepsilon_{\textrm{lin}}$ is useful as such an upper bound for a particular problem if and only if the known data from classical algorithms has asymptotic scaling that is measurably superlinear at that value of $\varepsilon_{\textrm{lin}}$ for that problem.  One reason for such a proposal is that quantum speedup is possible in principle at a given $\varepsilon$ for a given problem only if the known scalings for classical algorithms at that same value of $\varepsilon$ for that problem are all superlinear (again because linear scaling is the best possible scaling); a useful value of $\varepsilon_{\textrm{lin}}$ is one that eliminates such a value of $\varepsilon$ from quantum-speedup candidacy.  Further, it is intuitive that if a single classical heuristic without fixed scaling is found to be superlinear at $\varepsilon_{\textrm{lin}}$, then it is likely (though not guaranteed) that the other classical heuristics without fixed scaling are also superlinear there.

We note that there are very many classical heuristics for approximate optimization that do not have fixed scaling, and it would be beyond the scope of the present work to test all of them at a computed $\varepsilon_{\textrm{lin}}$ for a given problem; we suffice below with testing only two of the most popular ones.  

\begin{figure}
\includegraphics[scale=0.8]{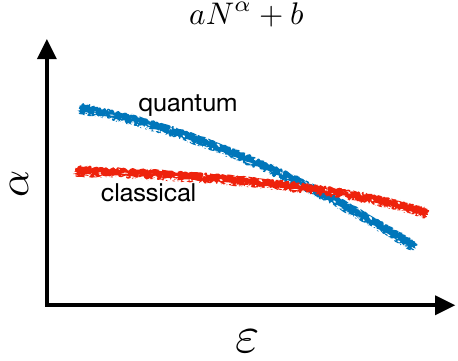}
\caption{(color online). Hypothetical example of polynomial scaling exponent ($\alpha$) vs. optimality gap for quantum and classical approximate optimization algorithms.  As illustrated, in some cases $\alpha$ of a quantum algorithm compared to a classical algorithm can be inferior at smaller optimality gaps ($\varepsilon$) but superior at larger optimality gaps.  This was found to be the case in the experiments in Ref. \cite{bauza2025scaling}, for example.  Larger values of $\varepsilon$ can therefore be relevant in the search for quantum speedup, and ruling them out with a fixed-scaling heuristic is therefore useful.
}
\label{fig:scaling_quantum_vs_classical}
\end{figure}

We also note that while it may seem that larger values of $\varepsilon$ are less useful to eliminate than smaller values of $\varepsilon$, it can actually turn out to be the case that a quantum algorithm for a given problem exhibits a scaling advantage over a classical algorithm at larger values of $\varepsilon$ but not at smaller values of $\varepsilon$; see Fig. \ref{fig:scaling_quantum_vs_classical} for an illustration. Ref. \cite{bauza2025scaling}, for example, found this to be the case for quantum annealing compared to parallel tempering with isoenergetic cluster moves in the setting of a quasi-2D classical spin glass.

Let us take an example for the utility of $\varepsilon_{\textrm{lin}}$.  Refs. \cite{bauza2025scaling, pawlowski2025closing} present results for the scaling of two classical heuristics for the approximate optimization of a particular classical spin glass model.  The scaling results span an optimality-gap range of $0.005\leq\varepsilon\leq0.0125$.  In this range, the data from the classical heuristics shows superlinear scaling, and the range $0\leq\varepsilon\leq0.0125$ is therefore a valid optimality-gap search-space for quantum speedup according to that data.  If now a linear-time classical heuristic is run on the same problem and produces $\varepsilon_{\textrm{lin}}=0.009$, the optimality-gap search-space for quantum speedup on that problem is reduced to $0\leq\varepsilon<0.009$.

In what follows we find values of $\varepsilon_{\textrm{lin}}$ that are in the few-percent range for a subclass of Eq. (\ref{eqn:Ham}).  We note that it is not a priori obvious that this could ever be the case in practice.  Choosing a random configuration of spins, for example, is a classical heuristic that is linear-time by construction, but for problems of the type in Eq. (\ref{eqn:Ham}) it will likely produce a value of $\varepsilon_{\textrm{lin}}$ that is near 100\% (see Ref. \cite{bauza2025scaling} and its End Matter for an example).

The specific problem setting for this work is Eq. (\ref{eqn:Ham}) on the simple cubic lattice.  The state-of-the-art of D-Wave quantum annealing encompasses approximate optimization of such models: D-Wave has demonstrated the approximate optimization of a cubic-lattice classical Ising spin glass with over 5,000 spins with pure quantum annealing \cite{king2023quantum} to $\varepsilon\approx2\%$ and over 11,000 spins with hybrid quantum annealing to $\varepsilon<1\%$ \cite{raymond2023hybrid}.  Experimental investigation of the time complexity of quantum annealing, and therefore of quantum speedup, for approximate optimization in this problem setting is therefore possible.  Further, recent theoretical results \cite{bernaschi2024quantum, ghosh2024exponential, morawetz2025universal, finzgar2025counterdiabatic, hattori2025controlled, bargava2026} suggest the possibility of modifying quantum annealing so as to achieve a substantial improvement in the time complexity for this computational task compared to ordinary quantum annealing.  It is therefore timely to ask if \textit{useful} values of $\varepsilon_{\textrm{lin}}$ are possible in this problem setting.  In this work we provide proof-of-principle results as evidence that they are.

\subsection{A subsystem optimization-based meta-heuristic for cubic-lattice spin glasses}
\label{sec:metaheuristic}

The general idea of optimizing spin glasses via optimization of subsystems is not new \cite{dantzig1960decomposition, bertsekas2008nonlinear}.  Ref. \cite{zintchenko2015local}, for example, presents a hierarchical algorithm in which a full system is recursively divided into subsystems, and Ref. \cite{rosenberg2016building} demonstrates variants of a method that iterates over optimizing over subsets of spins while leaving the rest of the system fixed.  While the meta-heuristic presented in this work shares the subsytem-optimization spirit of the these works, it is different in the following crucial ways: (1) it aims at only \textit{approximate} optimization instead of exact optimization and (2) it has a guaranteed linear time-complexity.  

The classical meta-heuristic that we propose here involves dividing the cubic lattice into contiguous subsystems of equal size and optimizing those subsystems in a bootstrapped way with a classical heuristic (e.g. simulated annealing).  Since any classical optimization heuristic can be used for the subsystem optimizations, we refer to this scheme as a \textit{meta}-heuristic (just as in the case of Ref. \cite{rosenberg2016building}).  If the subsystem optimizer is stochastic (e.g. simulated annealing), it can be forced to run for a chosen fixed time, which would make the global heuristic linear-time by construction.  If the subsystem optimizer is based on tensor-network contraction, as detailed in Appendix \ref{sec:TN_details}, the time complexity of the subsystem optimizer is determined only by the subsystem's network geometry, which can be chosen to be the same for each subsystem of the cubic lattice, thereby again making the global heuristic linear-time by construction.  We are not aware of any other works where such a meta-heuristic is implemented in the cubic-lattice setting.

The feasibility of the meta-heuristic as a whole for approximate optimization can be understood via a very simple intuition about short-range-correlated spin glass Hamiltonians: bootstrapping of approximate \textit{local} energy minimization should lead to approximate \textit{global} energy minimization.  By ``short-range-correlated" we mean that the short-range Hamiltonian does not amount to an embedding of a long-range Hamiltonian (as in, for example, Ref. \cite{konz2021embedding}).

As mentioned, our meta-heuristic involves bootstrapping $\mathcal{O}(N)$ local subsystem optimizations.  While such a local optimization method will be biased against spin configurations with large-scale structures, such as droplets, it is intuitive that not all low-energy configurations will have such structures.  However, small perturbations to the bonds in classical cubic-lattice spin glasses can substantially alter the ground state \cite{krzkakala2005disorder, katzgraber2007temperature}, so it is not a priori obvious how small of an average error such a bootstrapping heuristic will yield, nor is it obvious how much fluctuation will be in that error from instance to instance within a fixed hardness class.  Our results from a wide range of problem instances of cubic-lattice tile planting reveal an upper bound on the average relative energy error of $\sim7.5\%$ and a variance that varies widely over the different instance classes.  The average-error upper bound is substantially smaller than the smallest upper bound of $11.8\%$ that is theoretically proven (assuming P$\neq$NP) for the error of non-heuristic polynomial-time approximation algorithms for Eq. (\ref{eqn:Ham}) (see End Matter of Ref. \cite{bauza2025scaling}).

\subsection{Outline for the rest of the work}

In Section \ref{sec:experiments} we present finite-size results for $\varepsilon_{\textrm{lin}}$ from a tensor-network implementation of our meta-heuristic for two classes of the cubic-lattice tile-planted-solution model across their full range of energy landscape ruggedness.  We also present scaling results from simulated annealing and parallel tempering at the most rugged instances of this model over a range of target $\varepsilon$ values to show that the asymptotic value of $\varepsilon_{\textrm{lin}}$ for those instances reduces the quantum-speedup $\varepsilon$-search-space for those instances.  We conclude in Section \ref{sec:summary} with a summary and outlook.

\section{Experiments}
\label{sec:experiments}
The main objective of this section is to demonstrate that values of $\varepsilon_{\textrm{lin}}$ can be computed that are useful in reducing the $\varepsilon$ search-space for quantum speedup in approximate optimization on the cubic lattice.  For this we use a tensor-network implementation of our subsystem optimization-based meta-heuristic (see Appendix \ref{sec:TN_details} for details) and the tile-planting model.  For instances of this model that exhibit the highest MCMC-hardness for exact optimization, we show that $\varepsilon_{\textrm{lin}}$ lies below a range of $\varepsilon$ over which both simulated annealing (SA) and parallel tempering with isoenergetic cluster moves (PT+ICM) exhibit superlinear scaling.

The cubic-lattice tile-planting model is characterized in Ref. \cite{hamze2018near}.  The model contains multiple base classes of instances; the three known as $F_{22}$, $F_{42}$, and $F_{6}$ can be generated by the Python library called Chook \cite{perera2020chook}. Following Ref. \cite{hamze2018near} we generate two families of instances: \texttt{gallus\_26} (a mixture of $F_{22}$ and $F_6$), and \texttt{gallus\_46} (a mixture of $F_{42}$ and $F_6$).  Both families are parameterized by $p_6$, which denotes the fraction of the full-system Hamiltonian that belongs to $F_6$.  For all instances we enable the option in the Chook library to scramble the ground states with gauge transformations.  The error is defined the same as in Eq. (\ref{eqn:error}); $E_{gs}$ is given exactly by the Chook library for each generated instance.

Between $p_6=0.8$ and $p_6=1$, the hardness of exact optimization with MCMC of \texttt{gallus\_26} and \texttt{gallus\_46} is shown in Ref. \cite{hamze2018near} to increase (roughly) monotonically with increasing $p_6$.  When $p_6=0.8$ the MCMC exact-optimization hardness of \texttt{gallus\_46} is equivalent to that of the $\pm J$ model, and the MCMC exact-optimization hardness of \texttt{gallus\_26} is slightly higher.  When $p_6=1$, \texttt{gallus\_26} and \texttt{gallus\_46} are equivalent and have an MCMC exact-optimization hardness a few orders of magnitude greater than the $\pm J$ model.  We mention this information about exact-optimization hardness only in passing; MCMC exact-optimization hardness may or may not reflect the MCMC-hardness to optimize to a particular non-zero value of $\varepsilon$.

For our implementation of our meta-heuristic, we use cubic fragments of size 5$\times$5$\times$5 since that is the largest that is computationally feasible; in Appendix \ref{sec:Appendix_B} we confirm that error decreases monotonically with increasing cubic-fragment size.  We use twenty problem instances for each data point.  To estimate the optimal value of $\beta$, we test the performance of our heuristic on instances from the $F_6$ class (i.e., $p_6=1$) with different values of $\beta$ and system sizes.  The results are in Fig. \ref{fig:F6_error_vs_L} in Appendix \ref{sec:appendix_error_vs_beta}, based upon which we choose $\beta=2$ for the rest of the simulations.  

\begin{figure}
\includegraphics[scale=0.53]{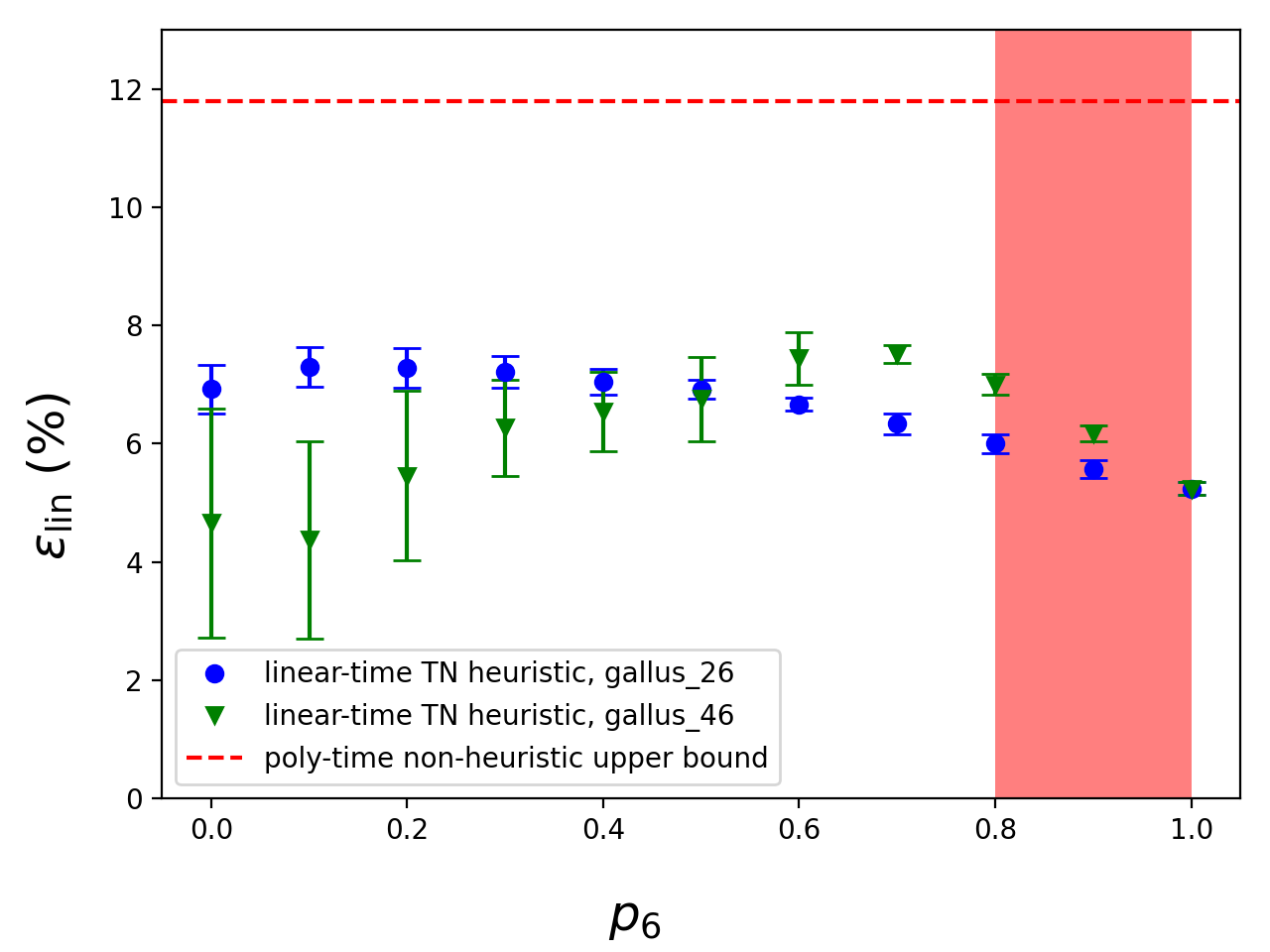}
\caption{(color online). Cubic-lattice tile-planting model (linear size $L=30$), classes \texttt{gallus\_26} (blue, discs) and \texttt{gallus\_46} (green, triangles), linear-time meta-heuristic ($l=5$ and $\beta=2$): optimality gap vs. $p_6$. Twenty instances at each $p_6$.  While Ref. \cite{hamze2018near} confirms monotonically increasing MCMC hardness (for exact optimization) with increasing $p_6$ between $p_6=0.8$ and $p_6=1$ (red shaded region), the error of our linear-time heuristic monotonically decreases.}
\label{fig:gallus_error}
\end{figure}

We first test our heuristic on \texttt{gallus\_26} and \texttt{gallus\_46} with $L=30$ and $\beta=2$ over many values of $p_6$.  The results are displayed in Fig. \ref{fig:gallus_error}.  We find an (empirical) upper bound on the error of about 7.5\%.  In passing, we note that optimality gaps of a few percent are acceptable in some types of industrial combinatorial optimization problems \cite{elyasi2024production}.  We also note that our error upper bound is almost a factor of two better than the upper bound of 11.8\% \cite{bauza2025scaling} known for non-heuristic approximate optimization algorithms for Eq. (\ref{eqn:Ham}) that operate in polynomial time.  However, this upper bound for non-heuristic approximate optimization takes into account the worst possible cases over all possible geometries for Eq. (\ref{eqn:Ham}).   Comparing to it the upper bound of our heuristic that is specialized to the cubic lattice is therefore not a completely fair comparison, but it is still of interest since it shows that our heuristic at least beats that more general upper bound of non-heuristic approximate algorithms.

Interestingly, though Ref. \cite{hamze2018near} confirms increasing landscape ruggedness (i.e., MCMC-hardness for exact optimization) as $p_6$ increases between 0.8 and 1, our linear-time heuristic actually shows a monotonically decreasing error.  As mentioned above, we do not know if MCMC exact-optimization-hardness correlates well with MCMC hardness at these values of $\varepsilon$, so it is not possible to use this by itself as good evidence that our method is immune to energy landscape ruggedness in configuration space. Nonetheless, we do know that our method does not perform a search over the energy landscape in configuration space, so it is possible that the value of $\varepsilon_{\textrm{lin}}$ obtained with our method is independent of landscape ruggedness; the data in Fig. \ref{fig:gallus_error} is consistent with that possibility.

\begin{figure}
\includegraphics[scale=0.5]{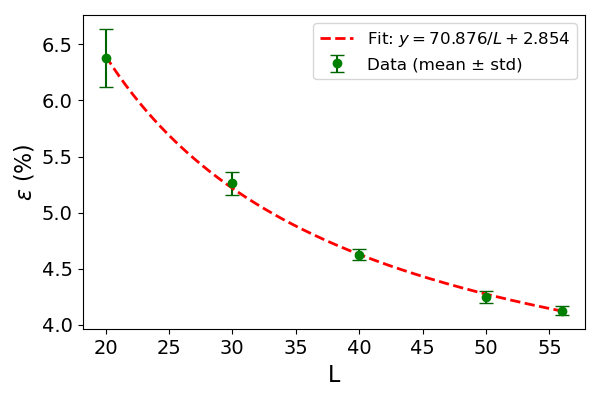}
\caption{(color online). Cubic-lattice tile-planting model, $F_6$ class (linear size $L$), linear-time meta-heuristic: optimality gap vs. linear system size ($L$) at $\beta=2$.  Data computed with cubic fragments of linear size $l=5$ spins.  Twenty instances at each $L$.  The fit assumes that the finite-size contribution to $\varepsilon_{\textrm{lin}}$ is due to the boundaries.}
\label{fig:TP3D_F6_error_beta_2}
\end{figure}

Next we investigate in more detail $p_6=1$ to demonstrate that the computed value of $\varepsilon_{\textrm{lin}}$ is useful there.  We expect the (periodic) boundaries to introduce error because there both ends of the fragment will overlap with decimated spins and the attempted energy minimization over the fragment will therefore be more constrained.  To estimate the asymptotic value of $\varepsilon_{\textrm{lin}}$, we fit a curve of the form $a/L + b$ to the data for $\varepsilon_{\textrm{lin}}$ over the range of $L$ that we test.  The fitting function is derived by considering that the boundaries cover a number of spins proportional to $L^2$, and that the finite-size error should decay as the ratio of boundary spins to bulk spins: $L^2/L^3=1/L$. Fig. \ref{fig:TP3D_F6_error_beta_2} reveals a very good fit, from which we take a conservative asymptotic optimality gap of $\varepsilon_{\textrm{lin}}=3\%$. To understand the scaling of simulated SA and PT+ICM as $\varepsilon_{\textrm{lin}}$ is approached from above, we therefore perform approximate optimizations with SA and PT+ICM to optimality-gap values of $\varepsilon=0.03, 0.04, $ and $0.05$ (see Appendix \ref{sec:appendix_SA_PT} for details about hyperparameter tuning).

\begin{figure}
\includegraphics[scale=0.55]{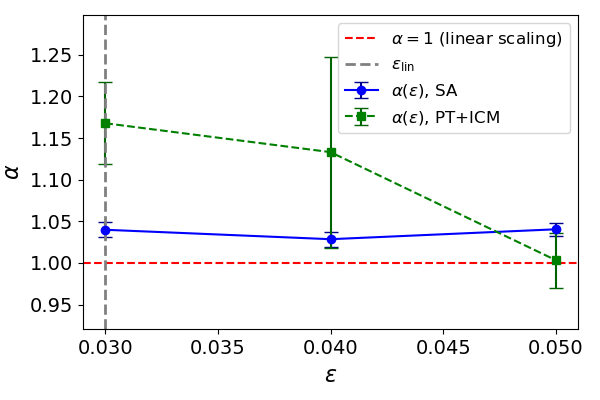}
\caption{(color online). Cubic-lattice tile-planting model ($L$$\times$$L$$\times$$L$ spins, periodic boundaries), $F_6$ class ($p_6=1$): polynomial scaling exponent ($\alpha$) vs. $\varepsilon$ for SA (blue, circles) and PT+ICM (green, squares). SA: twenty instances at each $L=20, 30, 40, 50, 56$; PT+ICM: twenty instances at each $L=6, 8, 10, 12$.  $\alpha(\varepsilon)$ computed by fitting to TT$\varepsilon$ = $aL^{3\alpha} + b$ at each $\varepsilon$. The values of $\varepsilon$ at which both SA and PT+ICM have $\alpha>1$ would be candidates for approximate-optimization quantum-speedup demonstration without the linear-time classical heuristic result, but they are ruled out for such quantum speedup by the computed value of $\varepsilon_{\textrm{lin}}$.}
\label{fig:TP3D_F6_SA_PT_scaling}
\end{figure}

Plotting the data for SA and PT on semilog plots vs. $N$ and several smaller powers of $N$ (not shown) does not reveal a straight line, so we conclude that the SA and PT scaling is most likely polynomial.  Fitting to $aN^{\alpha} + b$ for SA and PT+ICM reveals the $\alpha$ values shown in Fig. \ref{fig:TP3D_F6_SA_PT_scaling}.  It is clear that both SA and PT+ICM have superlinear scaling for a range of $\varepsilon>\varepsilon_{\textrm{lin}}$.  But since the definition of TT$\varepsilon$ is the time to reach down to or below an error of $\varepsilon$, we can say that the classical meta-heuristic has a TT$\varepsilon$ at  $\varepsilon\geq\varepsilon_{\textrm{lin}}$ that scales linearly.  Thus we conclude that in the $F_6$ class of problem instances, the computed value of $\varepsilon_{\textrm{lin}}$ is useful since it rules out the possibility of approximate-optimization quantum speedup at values of $\varepsilon$ where scaling data from SA and PT+ICM alone would allow it.

\section{Summary and Outlook}
\label{sec:summary}

The question of where quantum speedups lie for spin-glass optimization is an active area of research.  Establishing a quantum speedup for the approximate optimization of a specific problem is only possible at optimality gaps where no known classical algorithm has linear scaling.  A classical approximate optimization heuristic that has guaranteed linear scaling therefore produces an optimality gap that serves as an upper bound on the range of optimality gaps over which quantum speedup could occur. However, since linear time-complexity is the best possible scaling for any optimization algorithm, it is not a priori obvious that linear-time heuristics can produce optimality-gap upper bounds that are small enough to be useful; in this work we provide evidence that they can.

The cubic-lattice setting is relevant to the state-of-the-art of D-Wave's experimental capability for approximate optimization of classical spin glasses, and in this work we demonstrate a classical approximate optimization heuristic that is constructed with linear scaling and that therefore provides upper bounds on the optimality-gap search-space for quantum speedup on problems in that setting.  We provide evidence for the utility of such upper bounds by showing that two of the most popular classical heuristics, namely simulated annealing and parallel tempering with isoenergetic cluster moves, exhibit superlinear scaling at those same optimality-gap upper bounds.

The linear-time classical heuristic that we implement is actually one version of a meta-heuristic: instead of optimizing the full problem directly, it concatenates (via a bootstrapping procedure) the results of subsystem optimizations, but the subsystem optimizations can be done, in principle, with any heuristic.  We use a tensor-network heuristic for the subsystem optimizations.  The linear scaling of the meta-heuristic would not change by using a different heuristic for the subsystem optimizations.  Also of note is that certain choices of subsystem optimizers can have very few hyperparameters to tune, which makes for a very straightforward way of obtaining the optimality-gap upper bounds. In our tensor-network implementation of the meta-heuristic, for example, there is only one hyperparameter ($\beta$).

It is obviously desirable to obtain as low as possible of an upper bound on the optimality-gap search-space.  Though we do not expect a heuristic with \textit{linear} time-complexity to produce very low error solutions, there are a few ways that the meta-heuristic we present could be augmented to perhaps produce \textit{somewhat} lower error, and therefore lower such upper bounds.  One way is the following form of post-processing: undecimate the spins in a randomly selected fragment, approximately optimize that fragment again, accept the new configuration only if it lowers the global energy, and repeat on a new random fragment.  It would be straightforward to perform this post-processing in a way that preserves the linear scaling of the overall meta-heuristic, but one would still need to empirically verify that the error converges with increasing system size.  Also, we note that one of the possible sources of error in our implementation of the meta-heuristic, finite temperature, could be eliminated without changing the time complexity: the ordinary tensors could be replaced by tropical tensors \cite{liu2021tropical}.  This might reduce the global energy errors reported in this work at the same fragment sizes, but that is not guaranteed given the nature of spin-glass phases.  A third way to potentially lower the computed upper bounds is to try post-processing with randomized subsystem shapes and sizes while still enforcing the linear-time scaling.  Judging by Fig. \ref{fig:F6_error_vs_l}, a fourth way to lower the computed upper bound is to use larger fragment sizes; this will likely be technically prohibitive with exact tensor-network contraction as the subsystem optimizer, but it may be possible with other subsystem solvers such as simulated annealing.

By extension, fixed-scaling classical (meta-)heuristics can also lead to upper bounds for optimality-gap search spaces for quadratic and higher-order quantum speedups.  This is possible by constructing the time-complexity to be of the appropriate order.  To see this, recall the definition of quantum speedup ($S(N)$) from Ref. \cite{ronnow2014defining}:
\begin{equation}
S(N) = \frac{C(N)}{Q(N)},
\end{equation}
where $C(N)$ and $Q(N)$ are respectively the classical and quantum time-complexities under consideration.  For example, upper bounds for optimality-gap search-spaces for cubic quantum speedups in approximate optimization of classical spin glasses can be provided by constructing a classical (meta-)heuristic to have quartic time-complexity (i.e., $C(N)=\mathcal{O}(N^4)$) since the smallest that $Q(N)$ can be in principle is $\mathcal{O}(N)$.  We note that the significance of cubic quantum speedups is that only beyond-quadratic quantum speedups are expected to be practically relevant \cite{babbush2021focus}.

Industrially-relevant combinatorial optimization problems, when mapped onto spin glasses, tend to be native to non-local and dense graphs, and a subsystem-based meta-heuristic with fixed scaling could offer useful upper bounds on optimality gaps for quantum speedup in that setting.  There is already some work exploring subsystem-based meta-heuristics for non-local spin-glass optimization in Ref. \cite{rosenberg2016building} and Ref. \cite{cilasun20243sat}, but not in the sense of fixed scaling and upper bounds on quantum-speedup optimality gaps.  That could be another avenue for future work.

Finally, as mentioned, the subsystem optimizations in our meta-heuristic can be performed by any optimization heuristic.  We choose the optimization heuristic of tensor-network contraction as a proof-of-principle demonstration that it can work for our local-graph meta-heuristic.  We also expect to be able to perform such a proof-of-principle demonstration in the domain of subsystem-based meta-heuristics for non-local spin-glass optimization.  The significance of this is that, unlike simulated annealing and many other heuristics, the computational cost of tensor-network contraction is dominated by matrix multiplications, for which specialized photonic hardware is under active development.  Implementation of tensor-network contraction with such hardware may turn out to yield substantially lower time and energy costs than other subsystem-optimization methods.  However, such hardware is currently limited to 16-bit fixed-point numerical precision \cite{meng2025digital}, whereas the results presented here are all with 64-bit floating-point precision.  Our (unshown) preliminary tests of our tensor-network implementation of our meta-heuristic with \texttt{float16} showed very poor results for the cubic-lattice $\pm J$ model and cubic-lattice tile-planting model but modest results for Barahona's two-level spin glass \cite{barahona1982computational}, which is a cubic-lattice reduction of Max-Cut on random three-regular graphs.  Whether or not a tensor-network implementation of a subsystem-based meta-heuristic can be modified to generally yield sufficiently low errors on dense-graph problems when limited to 16-bit fixed-point precision is an open question.

\begin{acknowledgments}
We acknowledge discussions with Timothée Leleu, Sam Reifenstein, Victor Bastidas, Wangwei Lan, Johnnie Gray, Yu Tong, Tomislav Begušić, Garnet Chan, Salvatore Mandrà, and Sukhi Singh.  We acknowledge Humberto Munoz-Bauza for correspondence regarding the TAMC software package.
\end{acknowledgments}

\appendix

\section{Tensor-network implementation details}
\label{sec:TN_details}

\begin{figure*}
\includegraphics[scale=0.8]{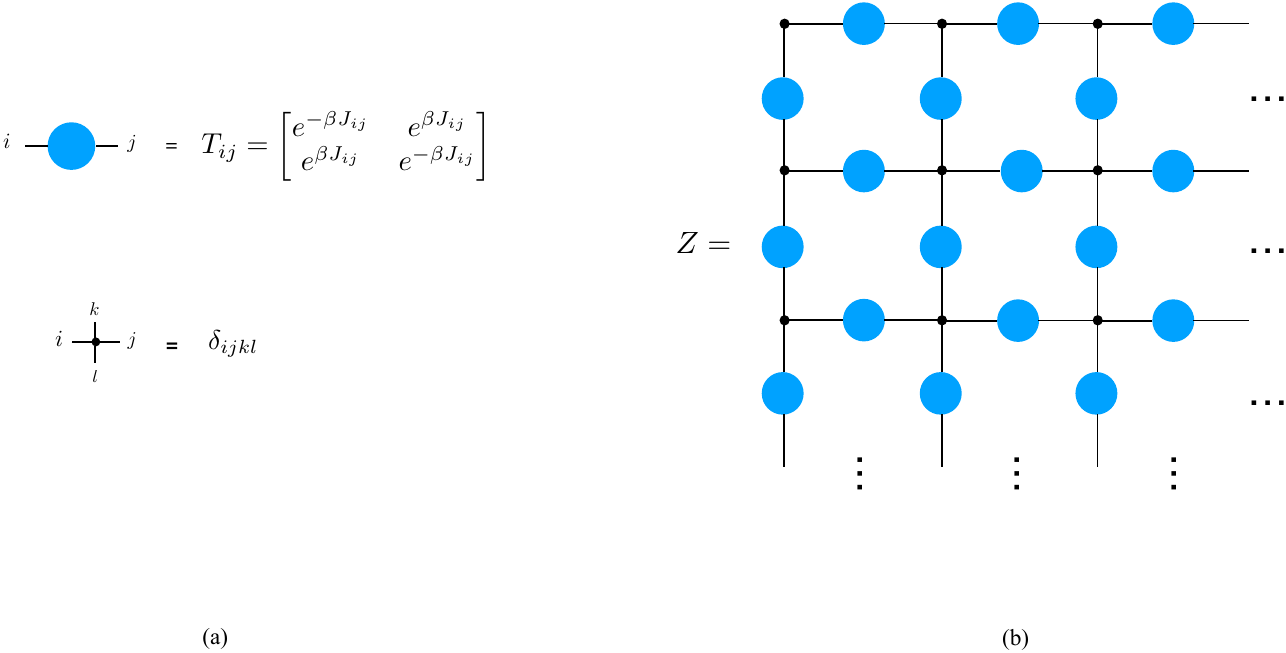}
\caption{(color online). (a) A large disc (blue) with two legs represents a two-index tensor of Boltzmann weights.  A small disc (black) with $n$ legs represents an $n$-index kronecker delta function.  The index dimensions are equal to the number of possible single-spin configurations (in this case two, corresponding to Ising spins). (b) Tensor network representation of the partition function for a square-lattice classical Ising model.  The delta functions are located at the sites of the spin lattice; they can alternatively be treated as hyperindices.  The joining of legs from different tensors represents a contraction of the tensors along a common (hyper) index.  The contraction of the entire network yields the partition function.
}
\label{fig:fig1}
\end{figure*}

\begin{figure*}
    \subfloat[]{{\includegraphics[width=0.35\textwidth]{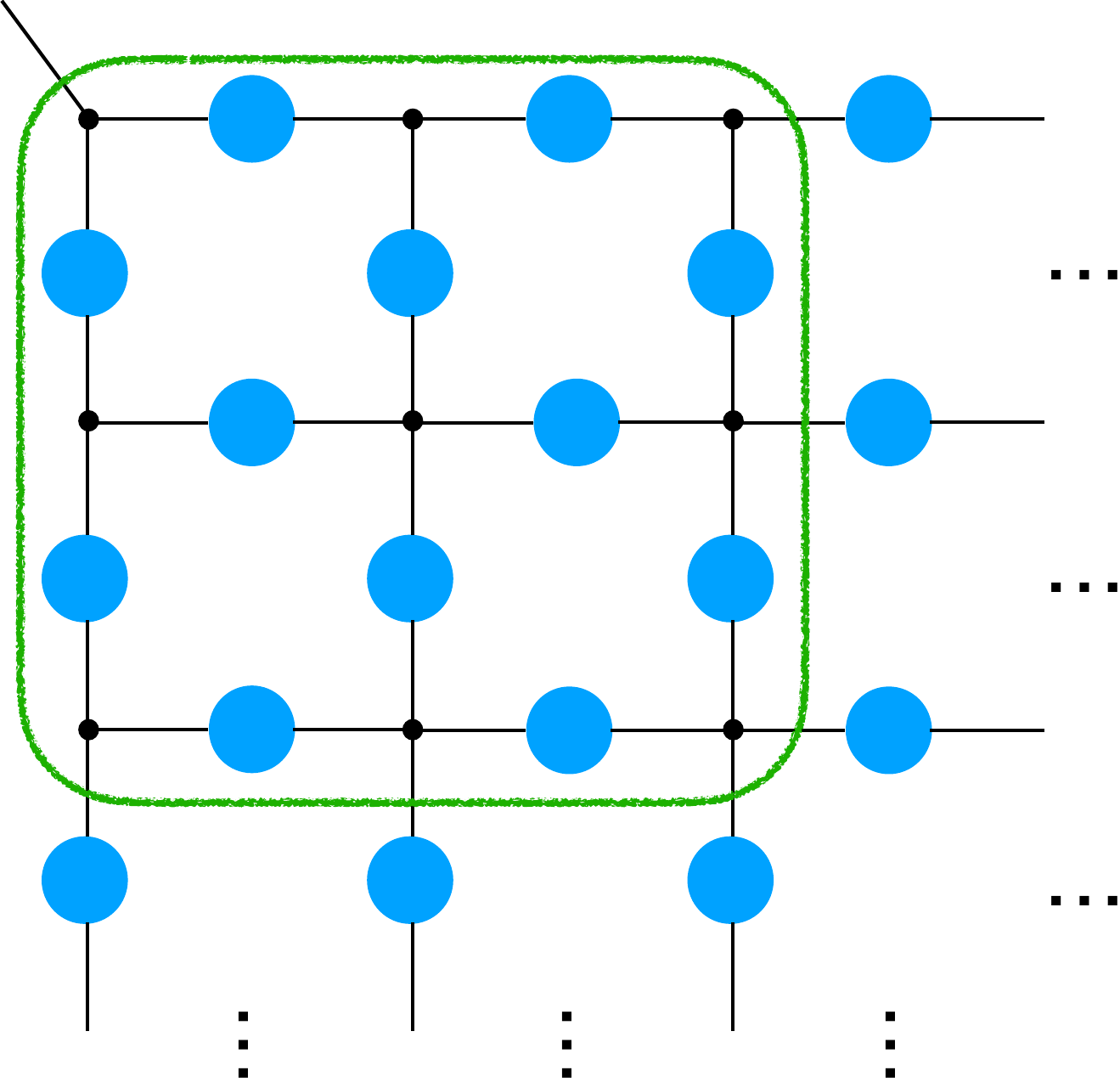} }} \hspace{3cm}
    \subfloat[]{{\includegraphics[width=0.35\textwidth]{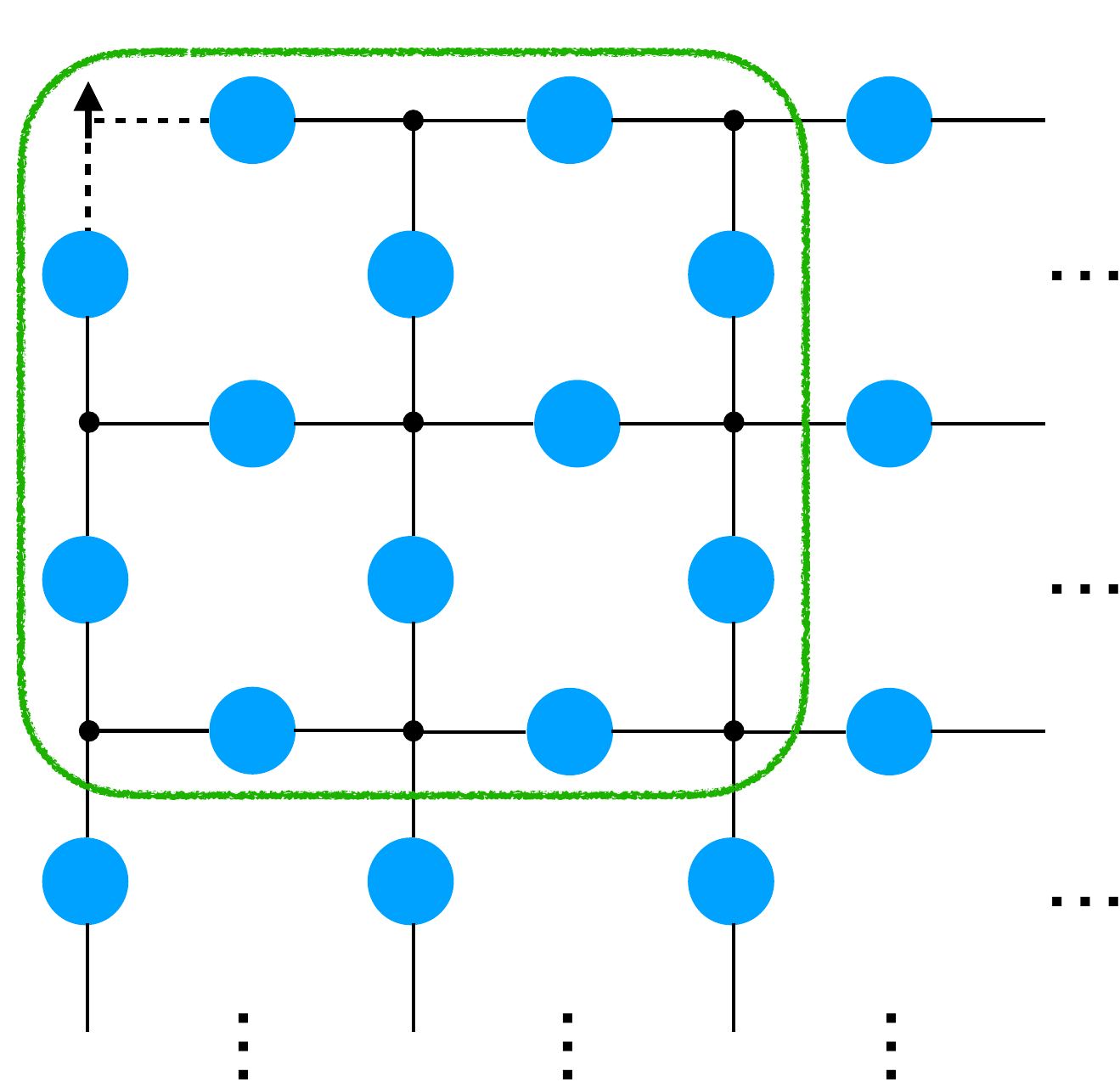} }}
    \vspace{1cm}
    
    \subfloat[]{{\includegraphics[width=0.35\textwidth]{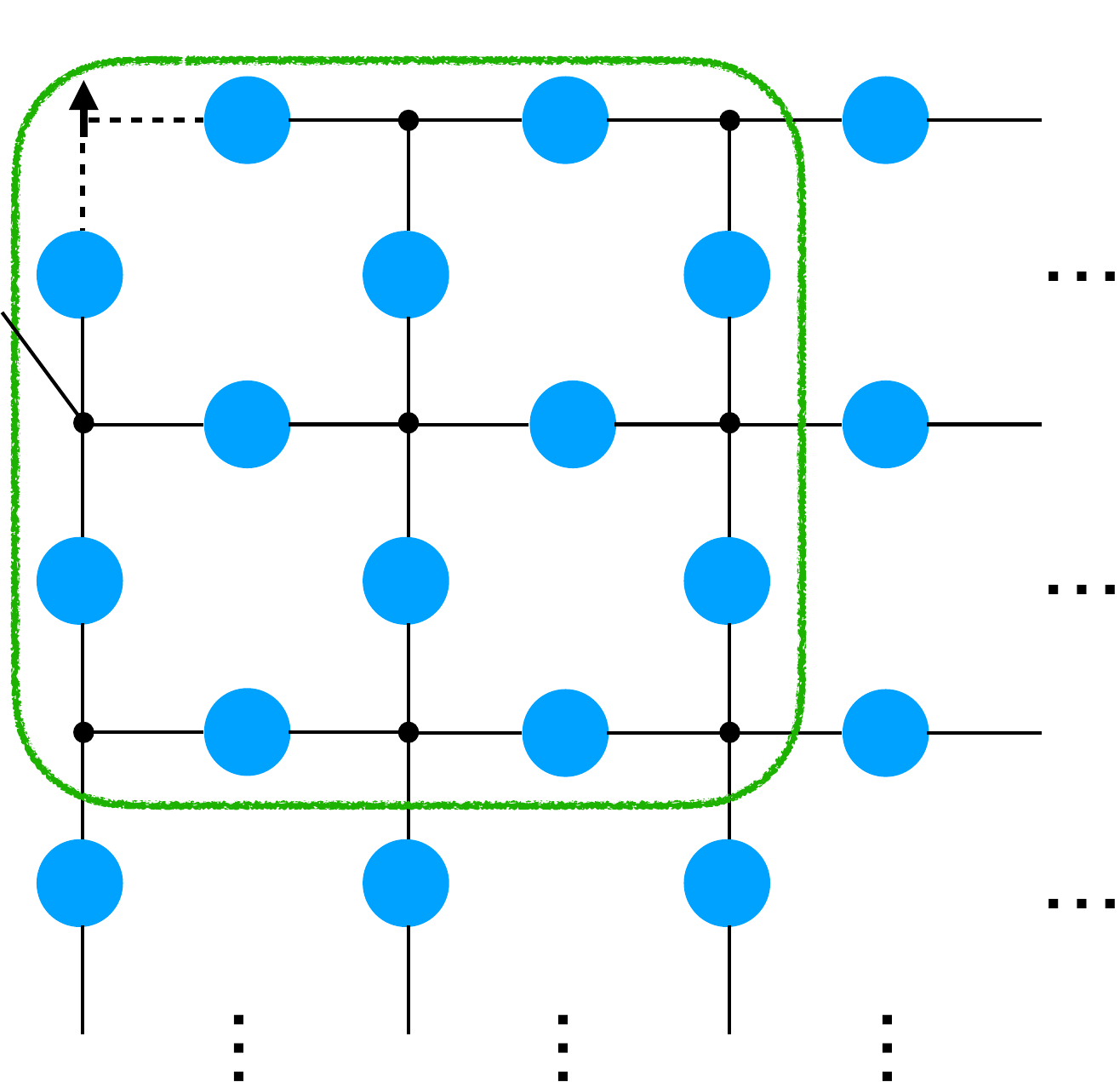} }} \hspace{3cm}
    \subfloat[]{{\includegraphics[width=0.35\textwidth]{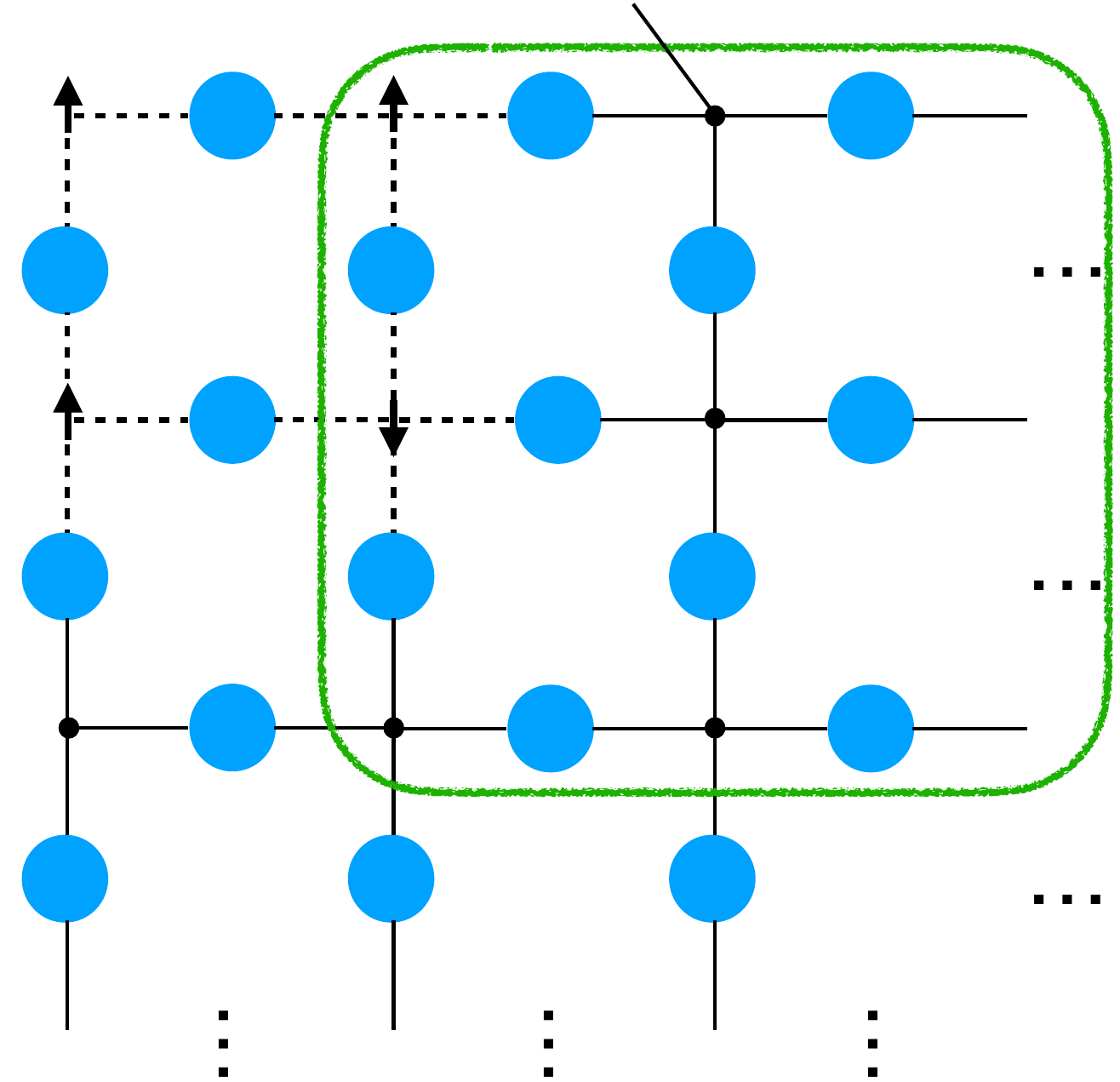} }}
    \vspace{1cm}
    
    \caption{(color online).  Local energy minimization bootstrapping procedure. (a) Adding an open leg to a single black disc turns it into a kronecker delta function such that the contraction of the entire network yields a vector that is the (unnormalized) unconditional marginal for the corresponding spin.  In the present algorithm, the approximation is made to  compute the marginals by contracting only a local fragment, defined by the tensors within the fuzzy (green) rounded-square boundary (tensors with legs that cross the boundary are not included in the fragment). The spin is decimated by choosing its most probable configuration according to this approximate marginal.  (b) The decimation is graphically denoted by legs with dashed lines and an up or down arrow (denoting $+1$ or $-1$).  The decimation is internally accomplished by selecting the appropriate value of the corresponding index.  (c) Adding an open leg to a different black disc after decimating previous spins yields the (unnormalized) marginal for the corresponding different spin.  The marginal for this spin is conditional upon the configuration of the previously decimated spins that lie within the fragment.  If $\beta$ is sufficiently large, decimating spins in this manner results in an approximate local energy minimization.  (d) Performing sequential decimations by overlapping the fragment with at least some of the previously decimated spins yields a bootstrapping of approximate local energy minimizations over the whole lattice.  Using multiple fragments simultaneously (not shown) yields parallelization.}
\label{fig:bootstrapping}
\end{figure*}

Our implementation of our classical meta-heuristic uses tensor-network contraction to approximately optimize the individual subsystems; this subsystem-optimization heuristic relies on a tensor-network representation of the partition function (see Fig. \ref{fig:fig1}), and is made possible through the following line of developments:  First, the work in Refs. \cite{nishino1995density, nishino1995product, nishino1998density, ueda2005snapshot} presented ways of using tensor-network algorithms for \textit{homogeneous} classical lattice models.  This included a method of sampling from the Boltzmann distribution of homogeneous, two-dimensional classical spin lattices \cite{ueda2005snapshot}.  Then, Ref. \cite{murg2005efficient} showed that the partition function of an \textit{inhomogeneous} classical spin lattice may be exactly represented by the contraction of a network of tensors where the geometry of the network reflects that of the Hamiltonian's interaction graph; while an exact contraction of the full network has an exponential cost, the contraction of the full network may be approximated, via truncated matrix decompositions, in polynomial time. Ref. \cite{rams2021approximate} then, analogously to the work in Ref. \cite{ueda2005snapshot} for homogeneous systems, demonstrated how to use this idea to sample (via computation of conditional marginals) the low-temperature Boltzmann distribution of planar and quasi-planar spin glasses in polynomial time.  Finally, Ref. \cite{gangat2024hyperoptimized} used a technically different type of approximate contraction from Ref. \cite{rams2021approximate} to generate near-optimal solutions of spin glasses on periodic square and cubic lattices in quadratic time.  We note, however, that the data in Ref. \cite{gangat2024hyperoptimized} for cubic-lattice spin glasses only spans two sizes (4$\times$4$\times$4 and 6$\times$6$\times$6) due to the high absolute time cost.

The subsystem-optimization heuristic that we use in our meta-heuristic differs from these previous tensor-network contraction methods in one crucial way: it utilizes only exact contractions of tensor-networks, which obviates the need for costly matrix decompositions.  This is feasible due to the limited sizes of the subsystems.

As explained in Refs. \cite{rams2021approximate, gangat2024hyperoptimized, chen2025tensor} both conditional and unconditional single-spin marginals for a classical spin glass in the canonical ensemble can be computed via the contraction of a tensor network wherein the indices of a given tensor correspond to single-spin configurations and the elements of the tensor correspond to the Boltzmann weights of the joint configurations of the spins at the tensor's legs (e.g., a two-index tensor contains the Boltzmann weights for all the possible configurations of two spins); the geometry of the tensor network mirrors the geometry of the Hamiltonian's interaction graph. The contraction of the network results in a multiplication of the local Boltzmann weights across the whole system that yields the (conditional or unconditional, depending on details of the network) marginal for the spin of interest.  See Fig. \ref{fig:fig1} and Appendix A of Ref. \cite{gangat2024hyperoptimized} for further explanation.  Computation of conditional single-spin marginals, and single-spin decimations according to those single-spin marginals, allows one to sample the Boltzmann distribution
\begin{equation}
    p(\textbf{s})\sim\textrm{exp}[-\beta H(\textbf{s})],
\end{equation}
where $\textbf{s}$ is a spin configuration vector for the entire system.  For computing the exact ground state, exact contractions of the network are required with the inverse temperature $\beta$ set to infinity.  While infinite $\beta$ is not numerically accessible, and exact contractions of the entire network are too costly for large systems, the intuition behind the heuristic in Ref. \cite{gangat2024hyperoptimized} is that \textit{approximate} contractions of the entire tensor network with sufficiently large $\beta$ should yield low-energy spin configurations.  For further details we refer the reader to Ref. \cite{gangat2024hyperoptimized}.

The tensor-network implementation of the meta-heuristic in this work can be seen as a modification of the method in Ref. \cite{gangat2024hyperoptimized}: instead of using approximate contractions of the entire tensor network, the marginal of a single spin is (approximately) computed by exactly contracting only a fragment of the network that is local to the spin of interest.  Sequentially computing single-spin marginals with tensor-network fragments that overlap previously decimated spins results in a bootstrapping of approximate local energy minimization if $\beta$ is sufficiently large (see Fig. \ref{fig:bootstrapping} for an illustration on the square lattice).  The intuition behind this approach is that if the Hamiltonian is only short-range correlated (see Sec. \ref{sec:metaheuristic} for definition of ``short-range correlated''), then such a bootstrapping should also result in an approximate \textit{global} energy minimization.  Such an outcome would be consistent with the finding in Ref. \cite{gangat2024hyperoptimized} that approximate optimization can be successfully accomplished for short-range-correlated Hamiltonians with approximate contractions of the full tensor network that use only small bond dimensions.

In this work we implement our tensor-network version of our meta-heuristic with the Python libraries \texttt{quimb} \cite{gray2018quimb} and \texttt{cotengra} \cite{gray2021hyper} on an Apple M2 Ultra CPU with 16 performance cores, 8 efficiency cores, and 128 GB of RAM.  We note that our meta-heuristic is straightforward to parallelize by using multiple fragments simultaneously, though we do not do so here.

\section{Solution quality vs. fragment size}
\label{sec:Appendix_B}
\begin{figure}
\includegraphics[scale=0.5]{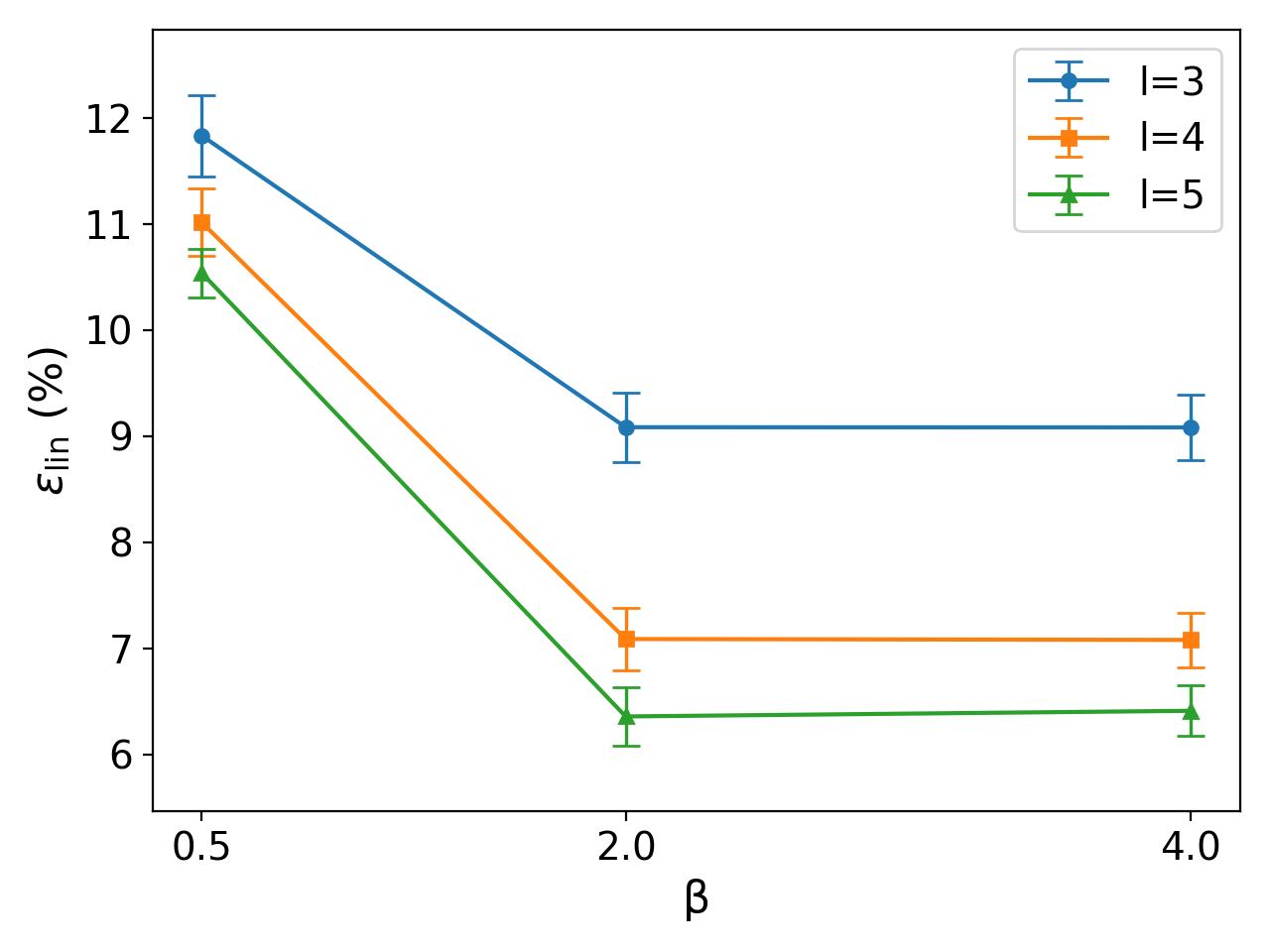}
\caption{(color online). Cubic-lattice tile-planting, $F_6$ class, $20$$\times$$20$$\times$$20$ spins, periodic boundaries, linear-time meta-heuristic: optimality gap vs. inverse temperature ($\beta$).  Data computed with cubic fragments of size $l$$\times$$l$$\times$$l$ spins on the same twenty instances at each $l$.}
\label{fig:F6_error_vs_l}
\end{figure}

Here we use the $F_6$ subclass of instances of the cubic-lattice tile-planting model, with system dimensions $20$$\times$$20$$\times$$20$ spins.  We test the solution quality of our method as a function of fragment size: we apply the bootstrapping algorithm to twenty instances of $L=20$ with cubic fragments of size $l$$\times$$l$$\times$$l$ spins with $l=3,4,$ and $5$; with $l=6$ we find the computation time of exactly contracting a single fragment to be impractically long.  The results are shown in Fig. \ref{fig:F6_error_vs_l}; as expected from the results in Ref. \cite{shen2025physics}, the error at a fixed $\beta$ decreases monotonically with $l$, so we use cubic fragments with $l=5$ for the computations in the main body.

\section{Solution quality vs. beta}
\label{sec:appendix_error_vs_beta}
\begin{figure}
\includegraphics[scale=0.5]{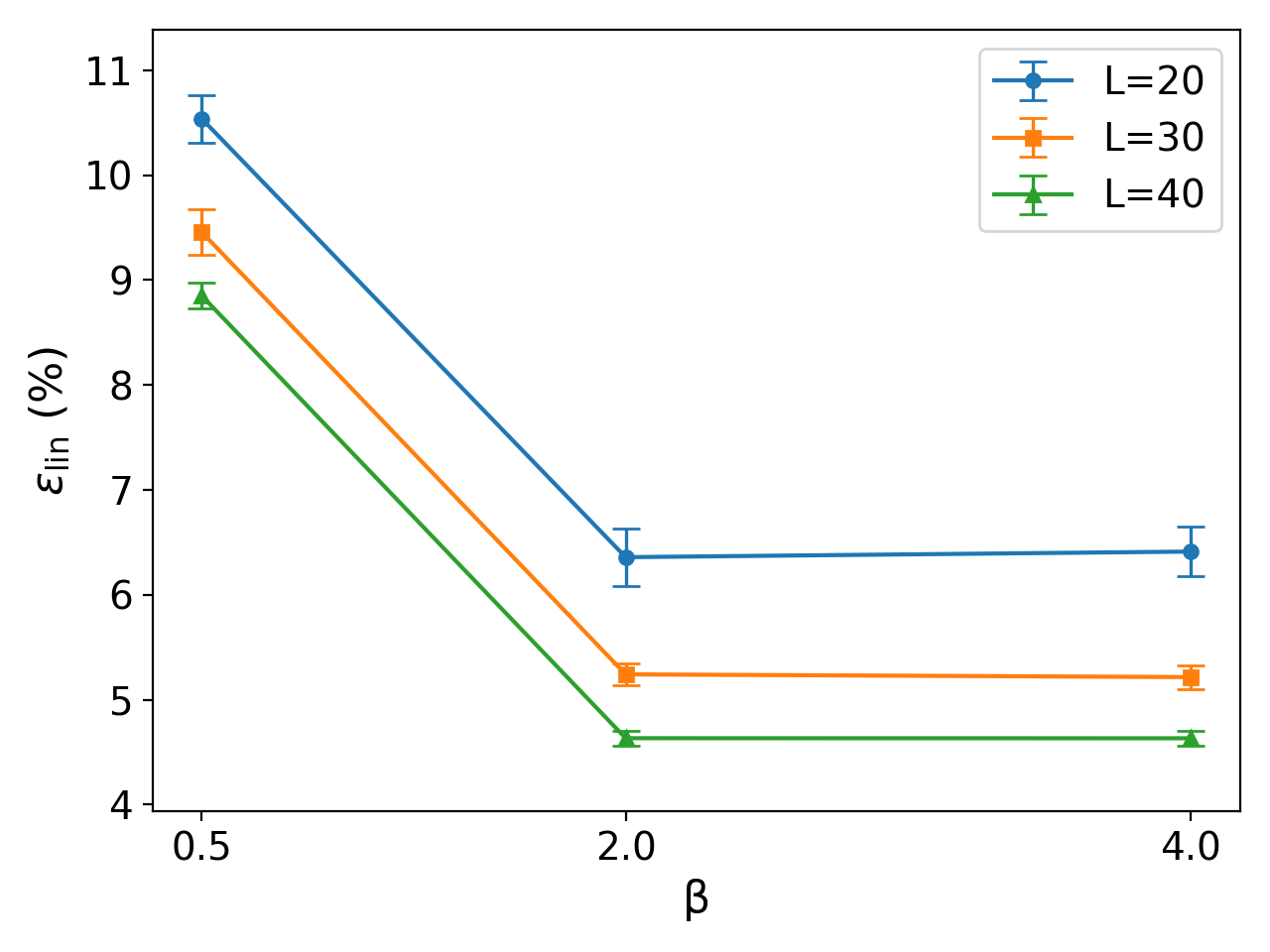}
\caption{(color online). Cubic-lattice tile-planting model, $F_6$ class (linear size $L$), linear-time meta-heuristic: optimality gap vs. inverse temperature ($\beta$).  Data computed with cubic fragments of linear size $l=5$ spins.  Twenty instances at each $L$.  Dashed lines connect data from same instances.}
\label{fig:F6_error_vs_L}
\end{figure}

The energy error of our heuristic (with $l=5$) vs. beta is shown in Fig. \ref{fig:F6_error_vs_L} for different $L$ averaged over twenty instances of the $F_6$ class (i.e., $p_6=1$) of cubic-lattice tile planting. The error is non-monotonic in $\beta$ due to finite numerical precision and possibly also finite fragment size, but monotonically decreases with increasing system size.  Based on this data, we choose $\beta=2$ for the simulations to compute $\varepsilon_{\textrm{lin}}$ for all the values of $p_6$ that we investigate in \texttt{gallus\_26} and \texttt{gallus\_46}.  Though the optimal value of $\beta$ may differ for different values of $p_6$, the choice of $\beta=2$ is sufficient for our purpose here of providing a proof of principle that $\varepsilon_{\textrm{lin}}$ can be small enough to be useful.

\section{Simulated Annealing and Parallel Tempering details}
The simulated annealing and parellel tempering simulations are run on an Apple M4 Max CPU with 12 performance cores, 4 efficiency cores, and 128 GB of RAM.  CPU-intensive background processes are terminated/disabled during the runs.

\label{sec:appendix_SA_PT}
\subsection{Simulated Annealing}

We use the simulated annealing implementation in the \texttt{dwave-samplers} library \cite{dwave}.  Metropolis sweeps are performed according to a geometric temperature schedule.  To minimize the TT$\varepsilon$, the number of reads is kept fixed at 1 since we find this to always be sufficient to reach the target $\varepsilon$ of interest.

We choose the temperature range for the annealing, independently for each target $\varepsilon$ and each problem class, by assessing the performance of different temperature ranges on problem instances at $L=10$:  First, for a given target $\varepsilon$ and problem class, for each of the 20 problem instances at $L=10$ we perform 20 independent runs of SA for each possible temperature range; the temperature ranges are from scanning the minimum $\beta$ between 0.3 and 3.3 in increments of 0.2 and the maximum $\beta$ between 3.0 and 27.0 in increments of 1.  The 20 independent runs at each possible temperature range are for accounting for the variation in performance arising from the stochasticity of simulated annealing as well as dependence on initial conditions.  The combination of minimum $\beta$ and maximum $\beta$ that minimizes the number of sweeps to reach the target $\varepsilon$ is recorded for each instance at $L=10$.  We consider the target $\varepsilon$ to be reached for a given temperature range and number of sweeps if the average $\varepsilon$ over 20 independent runs with that temperature range and number of sweeps is less than or equal to the target $\varepsilon$.  The average over all these instances of the optimal minimum $\beta$ and the optimal maximum $\beta$ are then used to fix the temperature range of the SA runs that are used to compute the scaling data for that problem class at that target $\varepsilon$.

Once the temperature ranges are tuned as mentioned, the smallest number of sweeps needed to reach the target $\varepsilon$ on average over 20 trials is found for a given problem instance, and the average walltime over those 20 trials at that minimum value of sweeps is used as the TT$\varepsilon$ for the given problem instance.  For a given $\varepsilon$ and $L$, the TT$\varepsilon$ is then averaged over the 20 problem instances at that $L$. These average values of TT$\varepsilon$ are then used to compute the scaling exponent for the given $\varepsilon$ by fitting to TT$\varepsilon$ = $aL^{3\alpha} + b$.

\subsection{Parallel Tempering}

We use the parallel tempering implementation in the \texttt{TAMC} software package \cite{tamc}.  We use a geometric temperature schedule.  The temperature range is tuned on instances at $L=4$, independently for each target $\varepsilon$, similarly as for SA at $L=10$, except that the minimum $\beta$ is scanned between 0.1 and 2.8 in increments of 0.2 and the maximum $\beta$ between 3 and 19 in increments of 1, and the optimal temperature range is considered to be the one that allows the target epsilon to be reached with the minimum of the product of the number of temperatures with the number of sweeps.  The number of sweeps is scanned between 1 and 20 in increments of 1, and the number of temperatures is scanned between 2 and 4 in increments of 1.  Isoenergetic cluster moves (ICM) are disabled for this temperature schedule tuning.

This temperature schedule tuning finds the optimal number of temperatures to be 2 for each value of $\varepsilon$ investigated (0.03, 0.04, and 0.05).  For the simulations with ICM enabled, we therefore set the number of cluster-move temperatures (\texttt{lo\_num\_beta}) to 2, which leads us to set \texttt{num\_replica\_chains=2}.

Once the hyperparameters are thusly tuned, they are kept fixed for computing the TT$\varepsilon$ at values of $L$ and target $\varepsilon$ of interest. The TT$\varepsilon$ is computed in the same manner as for simulated annealing.  We leave \texttt{warmup\_fraction=0.0} since we average over 20 trials for each instance to eliminate dependence on initial conditions. We set \texttt{threads=16}; the number of threads is chosen to be the maximum available on the version of the Apple M4 Max CPU that we use.

\newpage

\bibliography{Ref}

@article{mohseni2022ising,
  title={Ising machines as hardware solvers of combinatorial optimization problems},
  author={Mohseni, Naeimeh and McMahon, Peter L and Byrnes, Tim},
  journal={Nature Reviews Physics},
  volume={4},
  number={6},
  pages={363--379},
  year={2022},
  publisher={Nature Publishing Group UK London}
}

@article{abbas2024challenges,
  title={Challenges and opportunities in quantum optimization},
  author={Abbas, Amira and Ambainis, Andris and Augustino, Brandon and B{\"a}rtschi, Andreas and Buhrman, Harry and Coffrin, Carleton and Cortiana, Giorgio and Dunjko, Vedran and Egger, Daniel J and Elmegreen, Bruce G and others},
  journal={Nature Reviews Physics},
  pages={1--18},
  year={2024},
  publisher={Nature Publishing Group UK London}
}

@article{ronnow2014defining,
  title={Defining and detecting quantum speedup},
  author={R{\o}nnow, Troels F and Wang, Zhihui and Job, Joshua and Boixo, Sergio and Isakov, Sergei V and Wecker, David and Martinis, John M and Lidar, Daniel A and Troyer, Matthias},
  journal={Science},
  volume={345},
  number={6195},
  pages={420--424},
  year={2014},
  publisher={American Association for the Advancement of Science}
}

@article{farhi2001quantum,
  title={A quantum adiabatic evolution algorithm applied to random instances of an NP-complete problem},
  author={Farhi, Edward and Goldstone, Jeffrey and Gutmann, Sam and Lapan, Joshua and Lundgren, Andrew and Preda, Daniel},
  journal={Science},
  volume={292},
  number={5516},
  pages={472--475},
  year={2001},
  publisher={American Association for the Advancement of Science}
}

@article{venturelli2015quantum,
  title={Quantum optimization of fully connected spin glasses},
  author={Venturelli, Davide and Mandr{\`a}, Salvatore and Knysh, Sergey and O’Gorman, Bryan and Biswas, Rupak and Smelyanskiy, Vadim},
  journal={Physical Review X},
  volume={5},
  number={3},
  pages={031040},
  year={2015},
  publisher={APS}
}

@article{pirnay2024principle,
  title={An in-principle super-polynomial quantum advantage for approximating combinatorial optimization problems via computational learning theory},
  author={Pirnay, Niklas and Ulitzsch, Vincent and Wilde, Frederik and Eisert, Jens and Seifert, Jean-Pierre},
  journal={Science advances},
  volume={10},
  number={11},
  pages={eadj5170},
  year={2024},
  publisher={American Association for the Advancement of Science}
}

@article{jordan2025optimization,
  title={Optimization by decoded quantum interferometry},
  author={Jordan, Stephen P and Shutty, Noah and Wootters, Mary and Zalcman, Adam and Schmidhuber, Alexander and King, Robbie and Isakov, Sergei V and Khattar, Tanuj and Babbush, Ryan},
  journal={Nature},
  volume={646},
  number={8086},
  pages={831--836},
  year={2025},
  publisher={Nature Publishing Group UK London}
}

@article{boulebnane2024solving,
  title={Solving boolean satisfiability problems with the quantum approximate optimization algorithm},
  author={Boulebnane, Sami and Montanaro, Ashley},
  journal={PRX Quantum},
  volume={5},
  number={3},
  pages={030348},
  year={2024},
  publisher={APS}
}

@article{shaydulin2024evidence,
  title={Evidence of scaling advantage for the quantum approximate optimization algorithm on a classically intractable problem},
  author={Shaydulin, Ruslan and Li, Changhao and Chakrabarti, Shouvanik and DeCross, Matthew and Herman, Dylan and Kumar, Niraj and Larson, Jeffrey and Lykov, Danylo and Minssen, Pierre and Sun, Yue and others},
  journal={Science Advances},
  volume={10},
  number={22},
  pages={eadm6761},
  year={2024},
  publisher={American Association for the Advancement of Science}
}

@article{bauza2025scaling,
  title={Scaling advantage in approximate optimization with quantum annealing},
  author={Munoz-Bauza, Humberto and Lidar, Daniel A},
  journal={Physical {R}eview {L}etters},
  volume={134},
  number={1},
  pages={160601},
  year={2025},
  publisher={APS}
}

@article{pawlowski2025closing,
  title={Closing the Quantum-Classical Scaling Gap in Approximate Optimization},
  author={Pawlowski, J and Tarasiuk, P and Tuziemski, J and Pawela, L and Gardas, B},
  journal={arXiv preprint arXiv:2505.22514},
  year={2025}
}

@article{goto2019combinatorial,
  title={Combinatorial optimization by simulating adiabatic bifurcations in nonlinear Hamiltonian systems},
  author={Goto, Hayato and Tatsumura, Kosuke and Dixon, Alexander R},
  journal={Science advances},
  volume={5},
  number={4},
  pages={eaav2372},
  year={2019},
  publisher={American Association for the Advancement of Science}
}

@article{goto2021high,
  title={High-performance combinatorial optimization based on classical mechanics},
  author={Goto, Hayato and Endo, Kotaro and Suzuki, Masaru and Sakai, Yoshisato and Kanao, Taro and Hamakawa, Yohei and Hidaka, Ryo and Yamasaki, Masaya and Tatsumura, Kosuke},
  journal={Science Advances},
  volume={7},
  number={6},
  pages={eabe7953},
  year={2021},
  publisher={American Association for the Advancement of Science}
}

@inbook{caracciolo,
    author = {Caracciolo, Sergio and Hartmann, Alexander and Kirkpatrick, Scott and Weigel, Martin},
    title = {Spin Glass Theory and Far Beyond: Replica Symmetry Breaking After 40 Years},
    publisher = {World Scientific},
    year = {2023},
    chapter = {1}
}

@misc{Wangwei,
  title={(private communication)},
  author={Wangwei Lan}
}

@article{barahona1982computational,
  title={On the computational complexity of {I}sing spin glass models},
  author={Barahona, Francisco},
  journal={Journal of Physics A: Mathematical and General},
  volume={15},
  number={10},
  pages={3241},
  year={1982},
  publisher={IOP Publishing}
}

@article{king2023quantum,
  title={Quantum critical dynamics in a 5,000-qubit programmable spin glass},
  author={King, Andrew D and Raymond, Jack and Lanting, Trevor and Harris, Richard and Zucca, Alex and Altomare, Fabio and Berkley, Andrew J and Boothby, Kelly and Ejtemaee, Sara and Enderud, Colin and others},
  journal={Nature},
  volume={617},
  number={7959},
  pages={61--66},
  year={2023},
  publisher={Nature Publishing Group UK London}
}

@article{raymond2023hybrid,
  title={Hybrid quantum annealing for larger-than-QPU lattice-structured problems},
  author={Raymond, Jack and Stevanovic, Radomir and Bernoudy, William and Boothby, Kelly and McGeoch, Catherine C and Berkley, Andrew J and Farr{\'e}, Pau and Pasvolsky, Joel and King, Andrew D},
  journal={ACM Transactions on Quantum Computing},
  volume={4},
  number={3},
  pages={1--30},
  year={2023},
  publisher={ACM New York, NY, USA}
}

@article{bernaschi2024quantum,
  title={The quantum transition of the two-dimensional {I}sing spin glass},
  author={Bernaschi, Massimo and Gonz{\'a}lez-Adalid Pemart{\'\i}n, Isidoro and Mart{\'\i}n-Mayor, V{\'\i}ctor and Parisi, Giorgio},
  journal={Nature},
  volume={631},
  number={8022},
  pages={749--754},
  year={2024},
  publisher={Nature Publishing Group UK London}
}

@article{ghosh2024exponential,
  title={Exponential speed-up of quantum annealing via n-local catalysts},
  author={Ghosh, Roopayan and Nutricati, Luca A and Feinstein, Natasha and Warburton, PA and Bose, Sougato},
  journal={arXiv preprint arXiv:2409.13029},
  year={2024}
}

@article{morawetz2025universal,
  title={Universal counterdiabatic driving in Krylov space},
  author={Morawetz, Stewart and Polkovnikov, Anatoli},
  journal={PRX Quantum},
  volume={6},
  number={4},
  pages={040320},
  year={2025},
  publisher={APS}
}

@article{finzgar2025counterdiabatic,
  title={Counterdiabatic Driving with Performance Guarantees},
  author={Finzgar, Jernej Rudi and Notarnicola, Simone and Cain, Madelyn and Mikhail, Mikhail D Lukin and Sels, Dries},
  journal={arXiv preprint arXiv:2503.01958},
  year={2025}
}

@article{hattori2025controlled,
  title={Controlled Diagonal Catalyst Improves the Efficiency of Quantum Annealing},
  author={Hattori, Tomohiro and Tanaka, Shu},
  journal={arXiv preprint arXiv:2503.15244},
  year={2025}
}

@article{konz2021embedding,
  title={Embedding overhead scaling of optimization problems in quantum annealing},
  author={K{\"o}nz, Mario S and Lechner, Wolfgang and Katzgraber, Helmut G and Troyer, Matthias},
  journal={PRX Quantum},
  volume={2},
  number={4},
  pages={040322},
  year={2021},
  publisher={APS}
}

@article{hamze2018near,
  title={From near to eternity: spin-glass planting, tiling puzzles, and constraint-satisfaction problems},
  author={Hamze, Firas and Jacob, Darryl C and Ochoa, Andrew J and Perera, Dilina and Wang, Wenlong and Katzgraber, Helmut G},
  journal={Physical Review E},
  volume={97},
  number={4},
  pages={043303},
  year={2018},
  publisher={APS}
}

@article{babbush2021focus,
  title={Focus beyond quadratic speedups for error-corrected quantum advantage},
  author={Babbush, Ryan and McClean, Jarrod R and Newman, Michael and Gidney, Craig and Boixo, Sergio and Neven, Hartmut},
  journal={PRX quantum},
  volume={2},
  number={1},
  pages={010103},
  year={2021},
  publisher={APS}
}

@article{elyasi2024production,
  title={Production planning with flexible manufacturing systems under demand uncertainty},
  author={Elyasi, Milad and Altan, Ba{\c{s}}ak and Ekici, Ali and {\"O}zener, Okan {\"O}rsan and Yan{\i}ko{\u{g}}lu, {\.I}hsan and Dolgui, Alexandre},
  journal={International Journal of Production Research},
  volume={62},
  number={1-2},
  pages={157--170},
  year={2024},
  publisher={Taylor \& Francis}
}

@article{nishino1995density,
  title={Density matrix renormalization group method for 2D classical models},
  author={Nishino, Tomotoshi},
  journal={Journal of the Physical Society of Japan},
  volume={64},
  number={10},
  pages={3598--3601},
  year={1995},
  publisher={The Physical Society of Japan}
}

@article{nishino1995product,
  title={Product wave function renormalization group},
  author={Nishino, Tomotoshi and Okunishi, Kouichi},
  journal={Journal of the Physical Society of Japan},
  volume={64},
  number={11},
  pages={4084--4087},
  year={1995},
  publisher={The Physical Society of Japan}
}

@article{nishino1998density,
  title={A density matrix algorithm for 3d classical models},
  author={Nishino, Tomotoshi and Okunishi, Kouichi},
  journal={Journal of the Physical Society of Japan},
  volume={67},
  number={9},
  pages={3066--3072},
  year={1998},
  publisher={The Physical Society of Japan}
}

@article{murg2005efficient,
  title={Efficient evaluation of partition functions of inhomogeneous many-body spin systems},
  author={Murg, Valentin and Verstraete, Frank and Cirac, J Ignacio},
  journal={Physical {R}eview {L}etters},
  volume={95},
  number={5},
  pages={057206},
  year={2005},
  publisher={APS}
}

@article{rams2021approximate,
  title={Approximate optimization, sampling, and spin-glass droplet discovery with tensor networks},
  author={Rams, Marek M and Mohseni, Masoud and Eppens, Daniel and Ja{\l}owiecki, Konrad and Gardas, Bart{\l}omiej},
  journal={Physical Review E},
  volume={104},
  number={2},
  pages={025308},
  year={2021},
  publisher={APS}
}

@article{gangat2024hyperoptimized,
  title={Hyperoptimized approximate contraction of tensor networks for rugged-energy-landscape spin glasses on periodic square and cubic lattices},
  author={Gangat, Adil A and Gray, Johnnie},
  journal={Physical Review E},
  volume={110},
  number={6},
  pages={065306},
  year={2024},
  publisher={APS}
}

@article{chen2025tensor,
  title={Tensor network Monte Carlo simulations for the two-dimensional random-bond Ising model},
  author={Chen, Tao and Guo, Erdong and Zhang, Wanzhou and Zhang, Pan and Deng, Youjin},
  journal={Physical Review B},
  volume={111},
  number={9},
  pages={094201},
  year={2025},
  publisher={APS}
}

@article{ueda2005snapshot,
  title={Snapshot observation for 2d classical lattice models by corner transfer matrix renormalization group},
  author={Ueda, Kouji and Otani, Ryota and Nishio, Yukinobu and Gendiar, Andrej and Nishino, Tomotoshi},
  journal={Journal of the Physical Society of Japan},
  volume={74},
  number={Suppl},
  pages={111--114},
  year={2005},
  publisher={The Physical Society of Japan}
}

@article{gray2018quimb,
  title={quimb: A python package for quantum information and many-body calculations},
  author={Gray, Johnnie},
  journal={Journal of Open Source Software},
  volume={3},
  number={29},
  pages={819},
  year={2018}
}

@article{gray2021hyper,
  title={Hyper-optimized tensor network contraction},
  author={Gray, Johnnie and Kourtis, Stefanos},
  journal={Quantum},
  volume={5},
  pages={410},
  year={2021},
  publisher={Verein zur F{\"o}rderung des Open Access Publizierens in den Quantenwissenschaften}
}

@article{dantzig1960decomposition,
  title={Decomposition principle for linear programs},
  author={Dantzig, George B and Wolfe, Philip},
  journal={Operations research},
  volume={8},
  number={1},
  pages={101--111},
  year={1960},
  publisher={INFORMS}
}

@article{sciorilli2025competitive,
  title={A competitive NISQ and qubit-efficient solver for the LABS problem},
  author={Sciorilli, Marco and Camilo, Giancarlo and Maciel, Thiago O and Canabarro, Askery and Borges, Lucas and Aolita, Leandro},
  journal={arXiv preprint arXiv:2506.17391},
  year={2025}
}

@article{bertsekas2008nonlinear,
  title={Nonlinear Programming, Athena Scientific, Belmont, MA, 1999},
  author={Bertsekas, DP},
  journal={Cited Here},
  year={2008}
}

@article{zintchenko2015local,
  title={From local to global ground states in {I}sing spin glasses},
  author={Zintchenko, Ilia and Hastings, Matthew B and Troyer, Matthias},
  journal={Physical Review B},
  volume={91},
  number={2},
  pages={024201},
  year={2015},
  publisher={APS}
}

@article{rosenberg2016building,
  title={Building an iterative heuristic solver for a quantum annealer},
  author={Rosenberg, Gili and Vazifeh, Mohammad and Woods, Brad and Haber, Eldad},
  journal={Computational Optimization and Applications},
  volume={65},
  number={3},
  pages={845--869},
  year={2016},
  publisher={Springer}
}

@article{shen2025physics,
  title={The Physics of Local Optimization in Complex Disordered Systems},
  author={Shen, Mutian and Ortiz, Gerardo and Dong, Zhiqiao and Weigel, Martin and Nussinov, Zohar},
  journal={arXiv preprint arXiv:2505.02927},
  year={2025}
}

@article{bargava2026,
  title={Constant Depth Digital-Analog Counterdiabatic Quantum Computing},
  author={Bhargava, Balaganchi and Kumar, Shubham and Visuri, Anne-Maria and Erdman, Paolo A. and Solano, Enrique and Hegade, Narendra N.},
  journal={arXiv preprint arXiv:2601.01154},
  year={2026}
}

@article{meng2025digital,
  title={Digital-analog hybrid matrix multiplication processor for optical neural networks},
  author={Meng, Xiansong and Kong, Deming and Kim, Kwangwoong and Li, Qiuchi and Dong, Po and Cox, Ingemar J and Lioma, Christina and Hu, Hao},
  journal={Nature Communications},
  volume={16},
  number={1},
  pages={7465},
  year={2025},
  publisher={Nature Publishing Group UK London}
}

@article{perera2020chook,
  title={Chook--A comprehensive suite for generating binary optimization problems with planted solutions},
  author={Perera, Dilina and Akpabio, Inimfon and Hamze, Firas and Mandra, Salvatore and Rose, Nathan and Aramon, Maliheh and Katzgraber, Helmut G},
  journal={arXiv preprint arXiv:2005.14344},
  year={2020}
}

@article{liu2021tropical,
  title={Tropical tensor network for ground states of spin glasses},
  author={Liu, Jin-Guo and Wang, Lei and Zhang, Pan},
  journal={Physical {R}eview {L}etters},
  volume={126},
  number={9},
  pages={090506},
  year={2021},
  publisher={APS}
}

@article{katzgraber2007temperature,
  title={Temperature and disorder chaos in three-dimensional Ising spin glasses},
  author={Katzgraber, Helmut G and Krz{a}ka{\l}a, Florent},
  journal={Physical {R}eview {L}etters},
  volume={98},
  number={1},
  pages={017201},
  year={2007},
  publisher={APS}
}

@article{krzkakala2005disorder,
  title={Disorder chaos in spin glasses},
  author={Krz{\k{a}}ka{\l}a, F and Bouchaud, J-P},
  journal={Europhysics Letters},
  volume={72},
  number={3},
  pages={472},
  year={2005},
  publisher={IOP Publishing}
}

@article{cilasun20243sat,
  title={3SAT on an all-to-all-connected CMOS Ising solver chip},
  author={C{\i}lasun, H{\"u}srev and Zeng, Ziqing and Kumar, Abhimanyu and Lo, Hao and Cho, William and Moy, William and Kim, Chris H and Karpuzcu, Ulya R and Sapatnekar, Sachin S},
  journal={Scientific reports},
  volume={14},
  number={1},
  pages={10757},
  year={2024},
  publisher={Nature Publishing Group UK London}
}

@misc{dwave,
  title={Ocean-{S}{D}{K}. https://docs.ocean.dwavesys.com/en
                /stable/index.html},
  author={D-Wave},
  year={2022}
}

@misc{tamc,
  title={T{A}{M}{C} software package. https://github.com/hmunozb/tamc},
  author={Munoz-Bauza, H},
  year={2023}
}

\end{document}